\documentclass[letter, 10pt, conference]{ieeeconf}      
\makeatletter
\def\endthebibliography{%
  \def\@noitemerr{\@latex@warning{Empty `thebibliography' environment}}%
  \endlist
}
\makeatother

\IEEEoverridecommandlockouts 
\let\labelindent\relax
\usepackage{bm}
\usepackage{enumitem}
\usepackage{commath}
\usepackage{graphicx}
\graphicspath{{figures/}}
\usepackage{amsmath}
\usepackage{subcaption}
\usepackage{breqn}
\usepackage[compress]{cite}
\usepackage[dvipsnames]{xcolor}
\usepackage{booktabs}
\usepackage{empheq}
\usepackage{amsfonts}
\usepackage{amssymb}

\usepackage{amsthm}
\usepackage{hyperref}
\usepackage{xcolor}
\usepackage{mathrsfs}
\usepackage{accents}
\usepackage{thmtools}
\usepackage{thm-restate}
\usepackage{float}
\usepackage{comment}
\usepackage{algpseudocode}
\usepackage{algorithm}

\theoremstyle{plain}

\newtheorem{lemma}{Lemma}
\newtheorem{definition}{Definition}

\newtheorem{property}{Property}

\newtheorem*{problem*}{Problem}
\newtheorem*{theorem*}{Theorem}
\newtheorem{assumption*}{Assumption}

\declaretheorem[name=Theorem]{thm}

\newcommand{\bluetext}[1]{{\color{black}#1}}

\newcommand{\myvar}[1]{\bm{#1}}

\newcommand{\barvar}[1]{\bar{{#1}}}

\newcommand{\myvardot}[1]{\dot{\myvar{#1}}}

\newcommand{\myset}[1]{\mathcal{#1}}
\newcommand{\mysetclose}[1]{\overline{{#1}}}
\newcommand{\mysetbar}[1]{\bar{\myset{#1}}}
\newcommand{\mysetbound}[1]{\partial \myset{#1}}

\newcommand{\Lip}[2]{\mathcal{L}_{#1}^{#2}}
\newcommand{\stepk}{{k\Delta t}}
\newcommand{\stepkplus}{{(k+1)\Delta t}}

\title{\LARGE \bf
 Differentiable Predictive Control with Safety Guarantees: A Control Barrier Function Approach
}

\author{ W. Shaw Cortez, J. Drgona, A. Tuor,  M. Halappanavar, and D. Vrabie
\thanks{\textsuperscript{\dag} W. Shaw Cortez, J. Drgona, A. Tuor, M. Halappanavar, and D. Vrabie are with the Data Science and Machine Intelligence Department at the Pacific Northwest National Laboratory, Richland, WA, USA. 
        E-Mail: {\tt\small $\{$w.shawcortez, jan.drgona, aaron.tuor, mahantesh.halappanavar, draguna.vrabie$\}$@pnnl.gov}
        }
\thanks{This research was supported by the U.S. Department of Energy, through the Office of Advanced Scientific Computing Research's “Data-Driven Decision Control for Complex Systems (DnC2S)” project. Pacific Northwest National Laboratory is operated by Battelle Memorial Institute for the U.S. Department of Energy under Contract No. DE-AC05-76RL01830.}
}

\makeatletter
\def\footnoterule{\relax%
  \kern-5pt
  \hbox to \columnwidth{\vrule width 0.5\columnwidth height 0.4pt\hfill}
  \kern4.6pt}
\makeatother

\begin{document}

\maketitle
\thispagestyle{empty}
\pagestyle{empty}

\begin{abstract}
We develop a novel form of differentiable predictive control (DPC)  with safety and robustness guarantees based on control barrier functions. DPC is an unsupervised learning-based method for obtaining approximate solutions to explicit model predictive control (MPC) problems. In DPC, the predictive control policy parametrized by a neural network is optimized offline via direct policy gradients obtained by automatic differentiation of the MPC problem. The proposed approach exploits a new form of sampled-data barrier function to enforce offline and online safety requirements in DPC settings while only interrupting the neural network-based controller near the boundary of the safe set. The effectiveness of the proposed approach is demonstrated in simulation.
\end{abstract}

\section{Introduction}

Learning-based controllers have shown promising results across many applications and are advantageous due to offline training, which alleviates the burden of online computation. A pitfall of learning-based control laws is their unpredictability during implementation. To address this, newer forms of learning-based controllers combine modern control methodologies, such as model predictive control (MPC), with learning-based approaches \cite{Hewing2020}. 
MPC is advantageous in providing stability and constraint satisfaction guarantees by repeatedly solving a receding horizon, optimal control problem at every time instant \cite{camacho2013model}. However, MPC requires significant computational resources for implementation.

Differentiable predictive control (DPC) is an unsupervised learning-based method for learning explicit neural control laws for model predictive control (MPC) problems\cite{Drgona2022,Drgona2021b}. 
DPC alleviates the computational burden of online MPC by learning a receding horizon controller offline. In DPC, the state and input constraints of MPC are incorporated into the DPC loss function in the form of penalty functions and aggregated with the MPC cost function. The control policy is a neural network trained offline via stochastic gradient descent with gradients obtained from automatic differentiation of the MPC problem cost functions and constraints. 
DPC has been implemented in several applications with high performance and low computational resources \cite{King2022,Drgona2021a}. There exist probabilistic guarantees of safety for DPC \cite{Drgona2022}, but there are no deterministic, robustness guarantees that DPC will always satisfy the constraints and stabilize the system. The problem considered here is to provide robust safety guarantees for DPC without requiring a sampling-based solution with a large computational burden. 

A promising avenue for ensuring safety is a (zeroing) control barrier function (ZCBF) \cite{ames2019}, which is a function that characterises a safe set and whose non-negativity of it's derivative on the constraint boundary ensures forward invariance of the set. Typically, the barrier function is implemented as a quadratic program (QP) at each time step. The QP-based controller aims to enforce the barrier condition, while staying minimally close to a given control law. Many new developments have addressed input constraints \cite{Ames2021}, multiple ZCBFs \cite{xu2018constrained, ShawCortez2022}, sampled-data control \cite{ShawCortez2021a,Breeden2022, Roque2022}, self-triggering \cite{Yang2019}, event-triggering \cite{Long2022},  input-to-state safety \cite{Alan2021}, adaptive/data-driven methods \cite{Lopez2021}, and high-order ZCBFs \cite{Tan2021a, xiao2019control}. 

In \cite{Roque2022}, a time-varying sampled-data barrier function was proposed with MPC, coined Corridor MPC, to provide continuous-time safety/robustness guarantees and remove the need for terminal constraints sets. That solution is promising for DPC due to the discrete-time implementation of DPC and the time-varying nature, which can be used for trajectory tracking applications. However that formulation along with the existing methods usually require guaranteeing the barrier function condition holds everywhere in the safe set. This can be highly restrictive and overly conservative. Furthermore, the conventional QP-formulation is solved at every time instant, which is inefficient, and it is unknown a priori when the barrier function will override the learning controller. Ideally, the barrier function would not interrupt the learning controller until absolutely necessary. The approach from \cite{ShawCortez2022} only implements the barrier function condition near the boundary, which is promising, but is only applicable to time-invariant safe sets and continuous-time controllers.

We propose a novel control strategy that couples DPC with a new sampled-data barrier function. \bluetext{One main contribution of this work is the development of a sampled-data barrier function that is less conservative than existing methods as it only requires the barrier condition to hold near the constraint boundary. Furthermore, the controller acts as an event-triggered control wherein the QP-based control is only implemented if the system state is sufficiently close to the constraint boundary. This promotes the notion that the learning-based controller should be implemented for its performance and only overridden if unsafe behavior is imminent. The second contribution is the combination of the sampled-data barrier function with DPC, both during training and online, to guarantee safety of the closed-loop system. The proposed controller is robust to perturbations and provides continuous-time safety guarantees. The proposed method is demonstrated in simulation. }

\subsubsection*{Notation}
In this paper, we use $\mathbb{W}$ for whole numbers, $\mathbb{R}$ for real numbers,  $\mathbb{R}_{>0}$ for all positive real numbers and  $\mathbb{R}_{\geq 0}$ for non-negative real numbers.  The composition operator $\alpha \circ \beta$ represents $\alpha(\beta)$. A continuous function, $\alpha: \mathbb{R} \to \mathbb{R}$ is a \textit{extended class-$\mathcal{K}$ function} if it is strictly increasing and $\alpha(0) = 0$. A continuous function, $\beta:\mathbb{R} \to \mathbb{R}$ is an \textit{class-$\mathcal{K}_\infty$ function} if it is strictly increasing, $\beta(0) = 0$ and $\lim_{r\to \infty} \beta(r) = \infty$. 

\section{Background}\label{ssec:SDZCBF}

Here we present the sampled-data barrier function from \cite{Roque2022}. Consider the nonlinear affine, perturbed system: 
\begin{equation}\label{eq:nonlinear affine perturbed}
 \myvardot{x} = \myvar{f}(\myvar{x}) + \myvar{g}(\myvar{x}) \myvar{u} + \myvar{w}(t),
\end{equation}
 where $\myvar{f}: \myset{X} \to \mathbb{R}^{n_x}$ and $\myvar{g}: \myset{X} \to \mathbb{R}^{n_x\times n_u}$ are locally Lipschitz continuous functions on their domain $\myset{X} \subseteq \mathbb{R}^{n_x}$, $\myvar{u}: \myset{X} \to \myset{U} \subseteq \mathbb{R}^{n_u}$ is the control input for a compact set $\myset{U}$, $\myvar{w}(t)\in \myset{W} \subset \mathbb{R}^{n_x}$ is a piecewise continuous disturbance on the compact set $\myset{W}$, and $\myvar{x}(t, \myvar{x}_0) \in \myset{X}$ is the state trajectory at $t$ starting at $\myvar{x}_0 \in \myset{X}$, which with abuse of notation we denote $\myvar{x}(t)$. 
 
 Let $h(\myvar{x},t): \myset{D} \times \mathbb{R} \to \mathbb{R}$, where $\myset{D}\subset \myset{X}$ is a compact set, be a continuously differentiable function with locally Lipschitz gradient on $\myset{D} \times \mathbb{R}$ and $\|\frac{\partial h}{\partial t}(\myvar{x}, t)\| \leq \bar{h}_t$, $\bar{h}_t \in \mathbb{R}_{\geq 0}$, for all $\myvar{x} \in \myset{D}, t \geq 0$. We define the safe set as:
\begin{equation}\label{eq:constraint set general}
\myset{C}(t) = \{\myvar{x} \in \mathbb{R}^{n_x}: h(\myvar{x},t) \geq 0\}
\end{equation}
and the boundary of the safe set as $\mysetbound{C}(t) = \{ \myvar{x} \in \mathbb{R}^{n_x}: h(\myvar{x}, t) = 0\}$.
 
 We define the following constants: $\bar{u} = \max_{\myvar{u} \in \myset{U}} \| \myvar{u}\|$ and $\bar{w} = \max_{\myvar{w} \in \myset{W}} \|\myvar{w} \|$. Let $\Lip{hf}{x}, \Lip{hg}{x}, \Lip{\alpha}{x}, \Lip{h t}{x} \in \mathbb{R}_{\geq 0}$ be the respective Lipschitz constants of $\frac{\partial }{\partial x} h(\myvar{x}, t) \myvar{f}(\myvar{x})$, $\frac{\partial }{\partial x} h(\myvar{x}, t) \myvar{g}(\myvar{x})$, $\alpha \circ h(\myvar{x}, t)$ for an extended class-$\mathcal{K}$ function $\alpha$, and $\frac{\partial}{\partial t}h(\myvar{x},t)$ w.r.t. $\myvar{x}$ on $\myset{D}$. The terms $\Lip{hf}{t}, \Lip{hg}{t}, \Lip{\alpha h}{t}, \Lip{ht}{t} \in \mathbb{R}_{\geq 0}$ are the respective Lipschitz constants of the above terms with respect to time. Finally, from continuity of $\myvar{f}$,  $\myvar{g}$, and $\frac{\partial }{\partial x} h$ we can upper bound the following terms: $\| \myvar{f}(\myvar{x}) + \myvar{g}(\myvar{x}) \myvar{u} \| \leq \chi $, $\chi \in \mathbb{R}_{\geq 0}$, for all $\myvar{x} \in \myset{D}$, $\myvar{u} \in \myset{U}$ and $\| \frac{\partial h}{\partial x} (\myvar{x},t) \| \leq \bar{h}_x$, $\bar{h}_x \in \mathbb{R}_{\geq 0}$ for all $\myvar{x} \in \myset{D}$, $t \geq 0$.

\begin{definition}[\cite{Roque2022}]\label{def:SDZCBF}
Consider the system \eqref{eq:nonlinear affine perturbed} and the compact sets $\myset{C}(t)$ and $\myset{D}$, where $\myset{C}(t)$ is defined by \eqref{eq:constraint set general} for a continuously differentiable function $h: \myset{D} \times \mathbb{R} \to \mathbb{R}$, and  $\myset{D} \supset \myset{C}(t)$ for all $t \geq 0$.
The function $h$ is a \emph{sampled-data zeroing control barrier function} (SD-ZCBF) if for a given $\Delta t >0$ there exists an extended class-$\mathcal{K}$ function $\alpha$, where $\alpha \circ h: \myset{D} \times \mathbb{R} \to \mathbb{R}$ is Lipschitz continuous on $\myset{D} \times \mathbb{R}$, such that for any point ${\myvar{x}} \in \myset{D}$ and $k \in \mathbb{W}$ there is a constant feedback input $\myvar{u}({\myvar{x}},\stepk) \in \myset{U}$ satisfying the following condition:
\begin{multline}\label{eq:sampling zcbf condition}
    \frac{\partial }{\partial x} h(\myvar{x}, \stepk) \left( \myvar{f}(\myvar{x})+ \myvar{g}(\myvar{x}) \myvar{u}(\myvar{x}, \stepk) \right)  + \alpha(h({\myvar{x}}, \stepk))  \geq \\ \nu \Delta t +\bar{h}_t + \bar{h}_x \bar{w},  
\end{multline}
for $\nu := (\Lip{\alpha h}{x} + \Lip{hf}{x} + \Lip{hg}{x} \bar{u}) ( \chi + \bar{w}) + (\Lip{\alpha h}{t} + \Lip{hf}{t} + \Lip{hg}{t} \bar{u}) $.

\end{definition}
\bluetext{
The following lemma is useful for ensuring safety of sampled-data systems:
\begin{lemma}[\cite{glotfelter2017nonsmooth}]\label{lem:glotfelter}
Let $\bar{\alpha}: \mathbb{R} \to \mathbb{R}$ be a locally Lipschitz, extended class-$\mathcal{K}$ function and $h:[0, t_1] \to \mathbb{R}$ be an absolutely continuous function. If $\dot{h}(t) \geq - \bar{\alpha}(h(t))$, for almost every $t \in [0, t_1]$, and $h(0) \geq 0$, then there exists a class-$\mathcal{KL}$ function $\beta: \mathbb{R}_{\geq 0} \times \mathbb{R}_{\geq 0} \to \mathbb{R}_{\geq 0}$ such that $h(t) \geq \beta(h(0), t)$, and $h(t) \geq 0$, $\forall t \in [0, t_1]$. 
\end{lemma}
}
The SD-ZCBF ensures continuous-time safety satisfaction for a discrete-time controller, i.e. a sampled-data system. The SD-ZCBF could be used in combination with MPC \cite{Roque2022}. However, the derivation of such a SD-ZCBF is non-trivial and conservative due to the requirement of checking \eqref{eq:sampling zcbf condition} over the entire region $\myset{D}$. The implementation of the SD-ZCBF and other barrier functions \cite{Roque2022, ames2019} is inefficient as it requires a QP to be solved at each instant in time. Furthermore, Definition \ref{def:SDZCBF} is restrictive in that it requires $\myset{D}$ to be compact, which precludes obstacle avoidance scenarios in which the safe set, $\myset{C}$, and thus $\myset{D}$ is unbounded. The objective here is to reduce the conservativeness and increase the efficiency of the SD-ZCBF to complement the fast implementation of DPC (see Section \ref{sec:DPC}) to develop a high-performing, yet provably safe solution.

\section{A Novel Sampled-Data Barrier Function}

In this section, we extend the capabilities of the SD-ZCBF. First, we make a minor extension of Lemma \ref{lem:glotfelter}:
\begin{property}\label{prop:typeII}
The  function, $\alpha: \mathbb{R} \to \mathbb{R}$ is locally Lipschitz continuous and the restriction of $\alpha$ to $\mathbb{R}_{\geq 0}$ is of class-$\mathcal{K}$.
\end{property}
\begin{lemma}\label{lem:nonsmooth comparison}
Let $\alpha: \mathbb{R}\to \mathbb{R}$ be a function satisfying Property \ref{prop:typeII} and $\psi:[0, t_1] \to \mathbb{R}$ be an absolutely continuous function. If $\dot{\psi}(t) \geq -\alpha(\psi(t))$ holds for almost every $t \in [0, t_1]$ for which $\psi(t) \in [-b, a]$, with $a,b \in \mathbb{R}_{>0}$, and $\psi(0) \geq 0$, then $\psi(t) \geq 0$ $\forall t \in [0, t_1]$. 
\end{lemma}
\begin{proof}
If $\psi(t) \leq a$ for all $t \in [0, t_1]$, then the proof follows from Lemma \ref{lem:glotfelter}. If for some time interval $\myset{I} \subset [0, t_1]$, $\psi(t) > a$ $\forall t \in \myset{I}$, then clearly $\psi(t) > a > 0$ on $\myset{I}$, and $\psi(t) \leq a $ on $[0, t_1] \setminus \myset{I}$. Without loss of generality suppose $[0, t_1] \setminus \myset{I}$ is a connected set. By assumption, $\dot{\psi}(t) \geq -\alpha(\psi(t))$ holds almost everywhere on $[0, t_1] \setminus \myset{I}$, and must also hold almost everywhere on $\mysetclose{[0, t_1] \setminus \myset{I}}$. The application of Lemma \ref{lem:glotfelter} to $\mysetclose{[0, t_1] \setminus \myset{I}}$ ensures that $\psi(t) \geq 0$ on $\mysetclose{[0, t_1] \setminus \myset{I}}$, which then completes the proof. Note that the proof of Lemma \ref{lem:glotfelter} requires $\alpha$ to be an extended class-$\mathcal{K}$ function, but the proof only uses the restriction of $\alpha$ to $\mathbb{R}_{\geq 0}$, which is satisfied via Property \ref{prop:typeII}.
\end{proof}

Next, we modify the SD-ZCBF formulation from Definition \ref{def:SDZCBF} as follows. Consider a continuously differentiable function $h: \myset{X} \times \mathbb{R} \to \mathbb{R}$, and function $\alpha$ satisfying Property \ref{prop:typeII}. Motivated by \cite{ShawCortez2022}, we define the following sets:
\begin{equation}\label{eq:annulus}
     \myset{A}(t)=\{ \myvar{x} \in \mathbb{R}^{n_x}:h(\myvar{x},t) \in [-b, a]\}
\end{equation} 
\begin{equation}\label{eq:annulus hocbf}
    \mysetbar{C} = \{\myvar{x} \in \myset{C}(t), \ \forall t \geq 0\},
    \mysetbar{A} = \{ \myvar{x} \in \myset{A}(t), \ \forall t \geq 0 \}, 
\end{equation}
for $a, b \in \mathbb{R}_{>0}$ and assuming $\mysetbar{A}$ is compact. The set $\myset{A}(t)$ can be thought of as an annulus that surrounds the boundary of $\myset{C}(t)$. The idea here is that we only need to ensure that $\dot{h} \geq -\alpha(h)$ holds on a region around the boundary, $\mysetbound{C}(t)$. 

Now for the computation of all bounds and Lipschitz constant terms in $\nu$, we can replace $\myset{D}$ with $\mysetbar{A}$, which reduces the conservatism in computing said bounds (to compare with the original SD-ZCBF formulation, we can define $\myset{D} = \mysetbar{C}\cup \mysetbar{A}$). Finally, to address sampling effects with respect to the new set $\myset{A}(t)$, we must ensure that any trajectory approaching the boundary  $\mysetbound{C}(t)$ is sampled inside $\myset{A}(t)$. To ensure this, we provide a minimum bound on $a$ as: 
\begin{align}\label{eq:min a}
a \hspace{0.1cm} > \bar{h}:= \hspace{0.1cm} & \underset{\myvar{x} \in \myset{X}, t\geq 0} {\text{max}}
\hspace{.3cm} h(\myvar{x}, t)  \\
& \text{s.t.} \hspace{.1cm}  \myvar{x} \in \myset{B}_{\eta \Delta t}(\myvar{y}), \forall \myvar{y} \in \mysetbound{C}(t)  \nonumber
\end{align}
for $\eta = \chi  + \bar{w}$ \bluetext{and $\myset{B}_{\eta \Delta t}(\myvar{y}) = \{ \myvar{x} \in \mathbb{R}^n: \|\myvar{y} - \myvar{x}\| \leq \eta \Delta t \}$}. We note that since $\mysetbar{A}$ is assumed to be compact, $\{\myvar{x} \in \mysetbound{C}(t),  \forall t \geq 0\}$ is also compact such that \eqref{eq:min a} is well-defined. The new sampled-data barrier function is defined as follows:
\begin{definition}\label{def:SDZCBF new}
Consider the system \eqref{eq:nonlinear affine perturbed} and a continuously differentiable function $h: \myset{X} \times \mathbb{R} \to \mathbb{R}$, with $\mysetbar{A}$ defined by \eqref{eq:annulus hocbf} and $\myset{A}(t)$ defined by \eqref{eq:annulus}, for which $\mysetbar{A}$ is compact. The function $h$ is a \emph{SD-ZCBFII} if i) for a given $\Delta t >0$,  \eqref{eq:min a} holds and ii) there exists a function $\alpha: \mathbb{R}\to \mathbb{R}$ satisfying Property \ref{prop:typeII} with $\alpha \circ h: \mysetbar{A} \to \mathbb{R}$  locally Lipschitz continuous on $\mysetbar{A}\times \mathbb{R}$, such that for any point $\myvar{x} \in \myset{A}(\stepk)$ for a given $k \in \mathbb{W}$ there is a constant feedback input $\myvar{u}_k \in \myset{U}$ satisfying the following condition:
\begin{equation}\label{eq:sampling zcbf condition new}
    \phi(\myvar{x}, \myvar{u}_k, \stepk)  \geq  \bar{\nu} \Delta t + \bar{h}_x \bar{w},  
\end{equation}
with $\phi(\myvar{x}, \myvar{u}, t)$ $:=$ $\frac{\partial }{\partial x} h({\myvar{x}}, t) \left( \myvar{f}(\myvar{x})+ \myvar{g}(\myvar{x}) \myvar{u} \right) +\frac{\partial }{\partial t} h(\myvar{x},t) + \alpha(h({\myvar{x}}, t)) $, and $\bar{\nu}$ $:=$ $\Big(\Lip{\alpha h}{x} + \Lip{hf}{x}$ + $\Lip{h t}{x}$ $+ \Lip{h g}{x} \bar{u} \Big) \eta$  $+ \Lip{\alpha h}{t} + \Lip{h f}{t} +\Lip{h t}{t} +  \Lip{h g}{t} \bar{u} $.
\end{definition}

We are now ready to state one of our main results:
\begin{thm}\label{thm:safety}
{For a given $\Delta t >0$,} suppose $h: \myset{X} \times \mathbb{R} \to \mathbb{R}$ is a SD-ZCBFII for the system \eqref{eq:nonlinear affine perturbed}. For any given $k \in \mathbb{W}$, suppose $\myvar{x}(\stepk) \in \myset{C}(\stepk)$. Consider the following cases: i) $\myset{D} = \mysetbar{C} \cup \mysetbar{A} \subset \myset{X}$ and ii) $\myset{D} \not\subset \myset{X}$. Consider \eqref{eq:nonlinear affine perturbed} in closed-loop with any $\myvar{u} = \myvar{u}(\myvar{x}(\stepk), \stepk) \in \myset{U} \ $ which is constant almost everywhere on $[\stepk,\stepkplus]$, and for which \eqref{eq:sampling zcbf condition new} holds on $\myset{A}(\stepk)$. 
Then 
\begin{enumerate}[label=\alph*)]
    \item If i) holds, then $\myvar{x}(t)\in \myset{C}(t)$ for all $t \in [k\Delta t, (k+1)\Delta t]$. If ii) holds and $\myset{B}_{\eta \Delta t}(\myvar{x}_k) \subset \myset{X}$, then $\myvar{x}(t)\in \myset{C}(t)$ for all $t \in [k\Delta t, (k+1)\Delta t]$.
    \item If i) holds and $\myvar{x}(0) \in \myset{C}(0)$ then $\myvar{x}(t) \in \myset{C}(t)$ for all $t \geq 0$. If ii) holds and $\myvar{u}(t)$ is such that $\myvar{x}(t) \in \myset{X}$ for all $t \geq 0$, then $\myvar{x}(t) \in \myset{C}(t)$ for all $t\geq 0$.
\end{enumerate} 
\end{thm}
\begin{proof}
a) For notation, let $\myvar{x}_k:= \myvar{x}(\stepk)$ and $\myvar{u}_k := \myvar{u}( \myvar{x}(\stepk), \stepk)$. First, for any $\myvar{x}(\stepk) \in \myset{C}$, the furthest a solution $\myvar{x}(t)$ can travel between sampling periods is bounded by $\| \myvar{x}(t) - \myvar{x}(t_0) \| \leq \int_{\stepk}^{\stepkplus} \| \myvar{f}(\myvar{x}(\tau)) + \myvar{g}(\myvar{x}(\tau)) \myvar{u}_k + \myvar{w}(\tau) \| d\tau  \leq \eta \Delta t$. We note that this bound holds even if the solution does not exist up until $\stepkplus$ as the solution will not have been able to travel to the maximum bound. If ii) holds, then by assumption the solution cannot leave the domain $\myset{X}$ over which local Lipschitz properties hold in the time interval $[\stepk, \stepkplus]$. Furthermore, by construction of $a$, if $\myvar{x}(\stepk) \in \myset{C}(\stepk) \setminus \myset{A}(\stepk)$, then $
\myvar{x}(t)$ cannot reach the set boundary, $\mysetbound{C}(t)$, within the sampling period. Thus before reaching the boundary of the safe set, any solution must be sampled in $\myset{A}(t)$. The next task is then to ensure that if $\myvar{x}_k \in \myset{A}(\stepk)$, then $\dot{h} \geq -\alpha(h)$ holds almost everywhere on $[\stepk, \stepkplus]$ so that the conditions of Lemma \ref{lem:nonsmooth comparison} hold.  

The remainder of the proof of part a) follows similar steps as that of Lemma 1 of \cite{Roque2022}, and is included for completeness. Due to the local Lipschitz nature of the dynamics and the almost everywhere constant control input, $\myvar{x}(t)$ is uniquely defined on $[\stepk, \tau_1]$ for some $\tau_1 > \stepk$ \cite[Thm. 54]{Sontag1998}. Furthermore, since $\myvar{x}_k$ is in the interior of $\mysetbar{A}$ and $\mysetbar{A}$ is compact, then by continuity of $\myvar{x}(t)$ there exists a $\tau_0 \in (\stepk, \tau_1]$ for which $\myvar{x}(t) \in \mysetbar{A}$ for all $t \in [\stepk,\tau_0]$. We define the following function on $[\stepk, \tau_0]$: $m{_k(t)} =  \phi(\myvar{x}(t), \myvar{u}_k, t)  - \phi(\myvar{x}_k, \myvar{u}_k, \stepk) $. Due to the standard properties of Lipschitz continuity and the fact that $\stepk < \tau_0 < \stepkplus \Rightarrow \tau_0 - \stepk < \Delta t$, it follows that:
\begin{multline}\label{eq:sampling_closeness_solutions}
||m_k(t)|| \leq  \Big( \Lip{\alpha h}{t}  + \Lip{h f}{t} +\Lip{h t}{t}+ \Lip{h g}{t} || \myvar{u}_k || \Big) \Delta t \\
 +\Big(\Lip{\alpha h}{x}  + \Lip{h f}{x}+ \Lip{h t}{x}+ \Lip{h g}{x}  || \myvar{u}_k ||  \Big)  || \myvar{x}(t) - \myvar{x}_k || 
\end{multline}

Next, we bound the change in state by: $\| \myvar{x}(t) - \myvar{x}_k \|  =  \int_{\stepk}^{\tau_0} \| \myvar{f}( \myvar{x}(t)) + \myvar{g}(\myvar{x}(t)) \myvar{u}_k + \myvar{w}(t)\| dt \leq \eta \Delta t$. Substitution into \eqref{eq:sampling_closeness_solutions} yields: $||m_k ( t)||  \leq \bar{\nu} \Delta t$.
 We add $ \frac{\partial }{\partial x}h(\myvar{x}(t), t) \Big(\myvar{f}(\myvar{x}(t)) + \myvar{g}(\myvar{x}(t)) \myvar{u}_k 
  + \myvar{w}(t) \Big) + \alpha(h(\myvar{x}(t), t) + \frac{\partial  }{\partial t}h(\myvar{x}(t),t)$ to both sides of \eqref{eq:sampling zcbf condition new} and substitute $m_{k(t)}$ which yields: $ \dot{h}(\myvar{x}(t),t)) + \alpha(h(\myvar{x}(t),t))  \geq  \bar{\nu}\Delta t + m_k(t) +
    \frac{\partial}{\partial x}h(\myvar{x}(t), t) \myvar{w}(t)  + \bar{h}_x \bar{w}$. From the fact that $ \|m_k( t)\| \leq \bar{\nu} \Delta t$ and $\| \frac{\partial}{\partial \myvar{x}}h(\myvar{x}, t) \myvar{w}(t) \| \leq \bar{h}_x \bar{w} $ it follows that  $\dot{h}(\myvar{x}(t), t) \geq -\alpha(h(\myvar{x}(t),t))$ almost everywhere on $[\stepk, \tau_0]$. Furthermore, since $\myvar{x}_k$ is sampled prior to reaching the boundary $\mysetbound{C}(t)$, we can define $\bar{a} = a - \bar{h}$, which by assumption is positive. Now we can state that $\dot{h}(\myvar{x}(t), t) \geq -\alpha(h(\myvar{x}(t),t))$ holds for all $t \in [\stepk, \tau_0]$ for which $h(\myvar{x}(t)) \in [-b, \bar{a}]$, such that  Lemma \ref{lem:nonsmooth comparison} ensures that $\myvar{x}(t)\in \myset{C}(t)$ for all $t \in [k\Delta t, \tau_0]$. The remainder of the proof for part a) follows by showing that the interval of existence of $\myvar{x}(t)$ extends to $[\stepk, \stepkplus]$ due to compactness of $\mysetbar{A}$, which we refer to the proof in Lemma 1 of \cite{Roque2022}.

b) Since $\myvar{x}(0) \in \myset{C}(0)$, $\myvar{x}(0) \in \myset{X}$, we can repeat the result of a) for every $k\in \mathbb{W}$ such that  $\myvar{x}(t) \in \myset{C}(t)$, $\forall t\in [\stepk,  \stepkplus]$, up until some $T>0$ at which point $\myvar{x}(t)$ leaves $\myset{X}$ and ceases to exist. If i) holds, then for every $k \in \mathbb{W}$, $\myvar{x}(\stepk) \in \myset{C}(\stepk) \subset \myset{X}$ and we can repeatedly apply a) for $k \to \infty$ such that $\myvar{x}(t) \in \myset{C}(t)$ for all $t \geq 0$. Similarly, if ii) holds and $\myvar{x}(t) \in \myset{X}$ for all $t\geq 0$, then $\myvar{x}(t)$ uniquely exists for all $t \geq 0$. Then we can repeatedly apply the result of a) for $k \to \infty$ since the local Lipschitz property of the dynamics still holds for all $t \geq 0$.
\end{proof}


\section{Safe Differentiable Predictive Control}\label{sec:DPC}

In this section we combine the DPC and SD-ZCBFII approaches to provide a safe, high performance control solution with deterministic guarantees that can be implemented in real-time. 
\begin{figure*}[ht!]
    \centering
    \includegraphics[width = \linewidth]{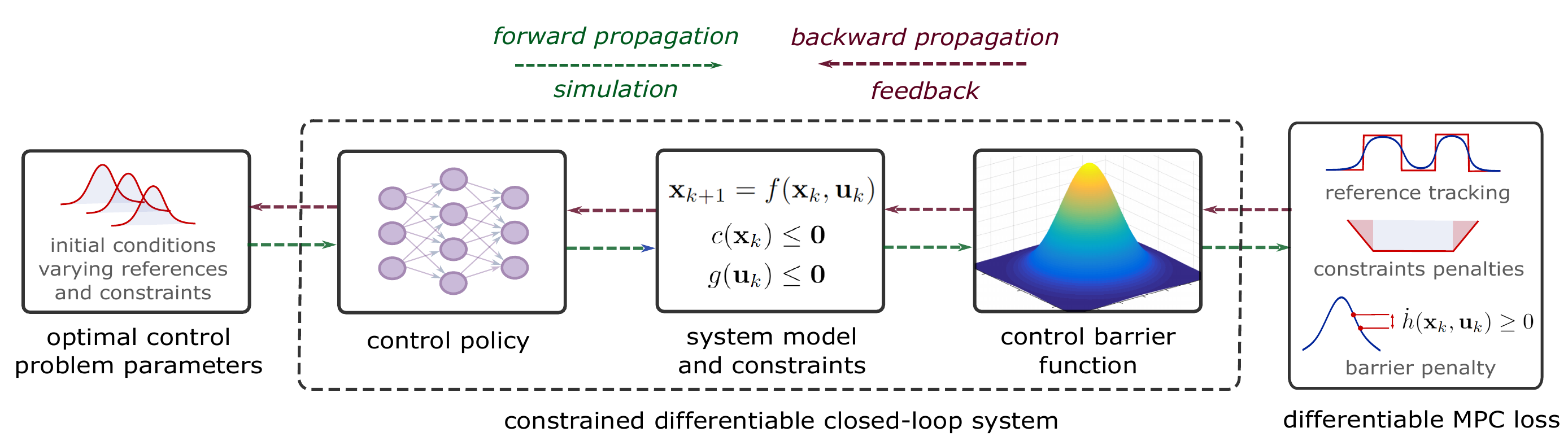}
    \caption{Conceptual methodology of training DPC offline with control barrier function constraints.}
    \label{fig:DPC_graph}
\end{figure*}
\subsection{Differentiable Predictive Control}\label{ssec:DPC intro}

DPC approximates a receding horizon control problem by solving the following DPC problem:
\begin{subequations}
\label{eq:DPC}
    \begin{align}
 \min_{\myvar{W}} & \frac{1}{mN} \sum_{i=1}^{m} \sum_{k=0}^{N-1}  \big( \ell_{\texttt{MPC}}( \myvar{x}_k^i, \myvar{u}_k^i,  \myvar{r}_k^i ) +  &  \label{eq:DPC:objective1}\\
 & p_x(c(\myvar{x}_k^i, \myvar{p}_{{c}_k}^i))  +   p_u( g(\myvar{u}_k^i, \myvar{p}_{{g}_k}^i))  \big) & 
 \label{eq:DPC:objective} \\ 
  \text{s.t.} \ &   \myvar{x}_{k+1}^i = A \myvar{x}_k^i + B \myvar{u}_k^i,    \label{eq:dpc:x}  & \\
 \  & \myvar{u}_k^i = \pi_{ {\bf W}}(\myvar{x}_k^i, \myvar{\xi}_k^i)  \label{eq:dpc:pi} \\
      \ &  \myvar{x}_0^i \in \mathbb{X} \subset \mathbb{R}^{n_{x}} \label{eq:dpc:x0} \\ 
  \ & \myvar{\xi}_k^i = \{ \myvar{r}_k^i,  \myvar{p}_{{c}_k}^i,  \myvar{p}_{{g}_k}^i\} \in \Xi \subset \mathbb{R}^{n_{\xi}} \label{eq:dpc:xi}
\end{align}
\end{subequations}
where the DPC loss function is composed of the parametric MPC objective $ \ell_{\texttt{MPC}}: \mathbb{R}^{n_x}\times \mathbb{R}^{n_u} \times \mathbb{R}^{n_{r}} \to \mathbb{R}  $, the parametrized state constraints $c:\mathbb{R}^{n_x} \times \mathbb{R}^{n_{p_c}} \to \mathbb{R}^{n_c}$,
the parametrized input constraints $g: \mathbb{R}^{n_u} \times \mathbb{R}^{n_{p_g}} \to \mathbb{R}^{n_g}$, and the penalty functions  $p_x: \mathbb{R}^{n_c} \to \mathbb{R} $, $ p_u: \mathbb{R}^{n_u}  \to \mathbb{R} $. In this setup, the model is a  discrete-time, linear approximation of the true system model \eqref{eq:nonlinear affine perturbed} defined in \eqref{eq:dpc:x}. The parameters of the problem are the components in \eqref{eq:dpc:xi}, which also treats the reference trajectory as a parameter to the problem. The parametric MPC loss objective can consider, for example, trajectory tracking error between the state and reference trajectory as well as input-related costs over a finite horizon. In addition to the loss function, the parametric constraints are treated as penalties to avoid control policies that violate constraints on the system. The MPC loss function, state constraints, input constraints, and penalty functions are assumed to be differentiable.

The formulation~\eqref{eq:DPC} implemented as a differentiable program allows us to obtain a data-driven, predictive solution
by differentiating the loss function \eqref{eq:DPC:objective1} and \eqref{eq:DPC:objective} backwards through the parametrized closed-loop dynamics given by the system model \eqref{eq:dpc:x}  and neural control policy \eqref{eq:dpc:pi}, which is parametrized by $\myvar{W}$. Stochastic gradient descent is used to minimize the loss function \eqref{eq:DPC:objective1} and \eqref{eq:DPC:objective}
over a distribution of control parameters \eqref{eq:dpc:xi} and initial conditions \eqref{eq:dpc:x0} sampled from the synthetically generated
training dataset $\Xi$, where $m$ represents the total number of parametric scenario samples, and $i$ denotes the index of the sample. The DPC problem \eqref{eq:DPC} is solved offline via Algorithm 2 of \cite{Drgona2022} to train the parameters of the control policy.


The advantage of DPC is its fast implementation online due to the offline computations of the control policy without requiring an expert controller for supervision, as is in the case of approximate MPC. The DPC setup is general, allowing for many different objectives, including, for example, trajectory tracking and obstacle avoidance. There are probabilistic guarantees that DPC will ensure constraint satisfaction and stability, but there is no deterministic guarantee of stability and safety, i.e., constraint satisfaction, nor robust guarantee to perturbations.

\subsection{Offline DPC Training with Barrier Functions}

In this section, we explain how the SD-ZCBFII can be incorporated into the training of the DPC. For a given SD-ZCBFII $h$, we approximate $\dot{h}(\myvar{x},t)$ using a finite-difference and define the following constraint to penalize any states that violate a barrier function-like condition:
\begin{multline}\label{eq:DPC barrier constraint}
    c_h(\myvar{x}_k^i, \myvar{x}_{k+1}^i, k, k+1) :=  h( \myvar{x}_{k+1}^i, k+1) - h(\myvar{x}_k^i, k) \\ + \alpha(h(\myvar{x}_k^i, k) )  - d
\end{multline}
where $d \in \mathbb{R}_{>0}$ is a robustness margin used to address the disturbance in the system. If $c_h \geq 0$, the barrier condition is satisfied. Otherwise, DPC will penalize the resulting policies that violate the condition. The condition \eqref{eq:DPC barrier constraint} is included in \eqref{eq:DPC} using a differentiable penalty function, $p_{ch}(c_h): \mathbb{R} \to \mathbb{R}$ and added to \eqref{eq:DPC:objective}. \bluetext{A graphical representation of the offline training approach is shown in Figure \ref{fig:DPC_graph}.}

The benefit of adding the barrier function condition into the training set is to promote control policies that attempt to minimize the MPC cost, while also satisfying the barrier condition. The intuition is that, in practice, the learned policy will avoid control inputs that violate the barrier function condition \eqref{eq:sampling zcbf condition new}.

\subsection{Online Safety Guarantees}

Suppose the DPC problem \eqref{eq:DPC} has been trained offline. The proposed control law is:
\begin{equation}\label{eq:safe DPC}
    \myvar{u}_k(\myvar{x}_k, \stepk) := \begin{cases}
    \myvar{\pi}_W(\myvar{x}_k, \stepk), \text{ if } \myvar{x}_k \notin \myset{A}(\stepk)  \\
    \myvar{u}^*_k(\myvar{x}_k, \stepk), \text{ if } \myvar{x}_k \in \myset{A}(\stepk)
    \end{cases}
\end{equation}
\begin{align}\label{eq:zcbf qp}
\myvar{u}_k^*(\myvar{x}_k, \stepk) \hspace{0.1cm} = \hspace{0.1cm} & \underset{\myvar{u} \in \myset{U}}{\text{argmin}}
\hspace{.3cm} \| \myvar{u} -\myvar{\pi}_{W}(\myvar{x}_k,\stepk) \|^2_2  \\
& \text{s.t.} \quad \phi(\myvar{x}_k, \myvar{u}, \stepk)  \geq  \bar{\nu} \Delta t +\bar{h}_x \bar{w}, \nonumber
\end{align}
 We note that $\phi(\myvar{x}_k, \myvar{u}, \stepk)$ is affine in $\myvar{u}$ such that the constraint in \eqref{eq:zcbf qp} is affine. Thus if $\myset{U}$ is a polyhedral set, $\eqref{eq:zcbf qp}$ is a QP. Furthermore, \eqref{eq:zcbf qp} need not be solved at all times. The optimization problem is only solved when the sampled state $\myvar{x}_k$ enters $\myset{A}(\stepk)$, which is more efficient than solving the QP at every instant in time. The proposed sampled-data formulation can be viewed as an event-triggered control law, wherein $\Delta t$ is the minimum dwell time and the triggering condition is $\myvar{x}_k \in \myset{A}(\stepk)$.

\begin{thm}\label{thm:safe dpc}
Suppose the conditions of Theorem \ref{thm:safety} hold and $\myvar{\pi}_W:\mathbb{R}^{n_x} \times \mathbb{R}\to \myset{U}$ is the control policy for which $\myvar{W}$ is the solution to \eqref{eq:DPC}. Consider the system \eqref{eq:nonlinear affine perturbed} in closed-loop with \eqref{eq:safe DPC}, \eqref{eq:zcbf qp}. If  $\myvar{x}(t) \in \myset{X}$ for all $t \geq 0$ or $\myset{D} \subset \myset{X}$, then $\myvar{x}(t) \in \myset{C}(t)$ for all $t \geq 0$.
\end{thm}
\begin{proof}
Since $h$ is a SD-ZCBFII, there always exists a solution to the control \eqref{eq:zcbf qp}. Furthermore, $\myvar{u}_k$ always satisfies $\myvar{u}_k \in \myset{U}$ and \eqref{eq:sampling zcbf condition new} is satisfied for any $\myvar{x}_k \in \myset{A}(\stepk)$. Thus the proof follows from Theorem \ref{thm:safety}. 
\end{proof}

One implementation of the SD-ZCBFII is as a `corridor', i.e., a tube around a desired reference trajectory to track. In this case, the SD-ZCBFII ensures boundedness of the state around a reference trajectory. Ideally the corridor would be made tight around the reference trajectory. Despite the improvements made to existing barrier functions, the SD-ZCBFII is still conservative particularly due to the fact that the sampling time is usually given and cannot be decreased to shrink the right-hand side of \eqref{eq:sampling zcbf condition new}. Herein lies the advantage of combining DPC with the SD-ZCBFII. By combining DPC with the SD-ZCBFII, we always guarantee robustness and safety, while keeping the desired performance of the system and reducing the computational resources required to compute the controller.

\section{Numerical Results}\label{sec:numerical results}

 We demonstrate our method on a simple, perturbed system: $\myvardot{x} = A\myvar{x} + \myvar{u} + \myvar{w}(t)$ for $\myvar{x}, \myvar{w}(t)\in \mathbb{R}^{n_x}$, and $\myvar{u} \in \myset{U} = \{ \myvar{u} \in \mathbb{R}^{n_x}: \|\myvar{u} \|_\infty \leq \bar{u} \}$. We aim to ensure the system tracks a desired reference trajectory with an ultimate-bound by using the SD-ZCBFII. The SD-ZCBFII is considered a `corridor' around the reference trajectory \cite{Roque2022} defined as \eqref{eq:constraint set general} with $ h(\myvar{x}, t) := \varepsilon - \|\myvar{x} - \myvar{x}_r(t) \|_2^2$ and for a given sufficiently smooth, time-varying reference $\myvar{x}_r: \mathbb{R} \to \mathbb{R}^{n_x} $ with bounds $\| \myvar{x}_r(t) \| \leq \bar{x}_r$, $\| \myvardot{x}_r(t) \| \leq \bar{v}_r$ for all $t \geq 0$ and $\bar{x}_r, \bar{v}_r \in \mathbb{R}_{>0}$. For the simulation, consider the following parameters: $n_x = 1$, $\varepsilon = 0.2$, $\myvar{w}(t) = 0.3sin(t)$, $\myvar{x}_r(t) = 0.5sin(0.5t)$, $A = 1.0$, $\bar{u} = 2$, and $\Delta t = 0.01s$. 
 
 We choose $\alpha(h) = \alpha h$ and construct $\barvar{u}_k$ to satisfy \eqref{eq:sampling zcbf condition new} as follows: $ \barvar{u}_k = \myvardot{x}_r(t) - A\myvar{x} - \frac{(\myvar{x} - \myvar{x}_r(t) )}{2 \|\myvar{x} - \myvar{x}_r(t)\|_2^2} \Big( \bar{\nu} \Delta t +\bar{h}_x \bar{w} - \alpha h(\myvar{x},t)  \Big)$. We note that $\barvar{u}_k$ blows up at $\myvar{x} = \myvar{x}_r(t)$. To avoid this, we choose $a < \varepsilon$ such that $\myvar{x} \neq \myvar{x}_r(t)$ for any $\myvar{x} \in \myset{A}(t)$. Since \eqref{eq:sampling zcbf condition new} is only required to hold in $\myset{A}(t)$, we can bound the following tracking error: $\|\myvar{x} - \myvar{x}_r(t)\|_\infty \geq \frac{1}{\sqrt{n_x}}\sqrt{\varepsilon - a}$ and $\|\myvar{x} - \myvar{x}_r(t)\|_\infty \leq \sqrt{\varepsilon + b}$. Note we use the equivalence of norms to determine these bounds as we will require the infinity-norm in $\myset{U}$. We take the norm of $\barvar{u}_k$ to check that $\barvar{u}_k\in\myset{U}$: $\|\barvar{u}_k\|_\infty \leq \bar{v}_r + \|A\|_\infty (\sqrt{c + b}+ \bar{x}_r)+ \frac{\sqrt{n_x}}{2 \sqrt{\varepsilon - a}}(\bar{\nu}\Delta t + \bar{h}_x \bar{w} + \alpha \max{\{a,b\}})$. The final condition that must hold is \eqref{eq:min a}, which can be checked using the following relations for $\myvar{y} \in \mysetbound{C}(t)$, $\myvar{x} \in \myset{B}_{\eta \Delta t}(\myvar{y})$: $| h(\myvar{x},t) - h(\myvar{y},t)| = \Lip{h}{x} \|\myvar{x} - \myvar{y}\| \leq \Lip{h}{x} \eta \Delta t$ and $h(\myvar{y}, t) = 0$ such that $|h (\myvar{x}, t)| \leq \Lip{h}{x} \eta \Delta t =: \bar{h}$, where $\Lip{h}{x}$ is the Lipschitz constant of $h(\myvar{x}, t)$ on $\mysetbar{A}\times \mathbb{R}$. We also note that here $\myset{X} \subseteq \mathbb{R}^{n_x}$ such that $\myset{D} = \mysetbar{C} \cup \mysetbar{A} \subset \myset{X}$, and so the conditions of Theorem \ref{thm:safe dpc} hold. For $\alpha = 0.5$, $ a = 0.03$, and $b = 0.00001$, $h$ is a SD-ZCBFII. 
 
 To compare with the original SD-ZCBF from Definition \ref{def:SDZCBF}, an additional step is required to show that there exist a control for $h > a$ to satisfy \eqref{eq:sampling zcbf condition} on all of $\myset{D}$. As done in \cite{Roque2022}, we need to check if $\barvar{u}_k = 0$ is a valid solution, which would require $\alpha$ to be tuned sufficiently large. However one would need $\alpha \geq 41$ to satisfy the condition, which in turn requires $\|\barvar{u}_k\| \leq 4.973 \not\leq \bar{u}$. Since the proposed control does not satisfy the input constraints, $h$ is \emph{not} a SD-ZCBF under these parameters, while it is a SD-ZCBFII. More design iterations would be required to see if $h$ is SD-ZCBF and most likely require larger control authority. This demonstrates the less conservative approach for the proposed method that also facilitates the barrier function design.

 The DPC Algorithm 2 of \cite{Drgona2022} is implemented using the
 Neuromancer framework~\cite{Neuromancer2021}. For training the neural control policies we assume the following unperturbed, discrete-time system approximation: $\myvar{x}_{k+1} = (A + \Delta t I) \myvar{x}_k + \Delta t \myvar{u}_k$. The barrier constraint \eqref{eq:DPC barrier constraint} for $d = 0.001$ was implemented along with the $log_{10}$ penalty functions. The DPC loss function used was: $\ell_{\texttt{MPC}}( \myvar{x}_k^i, \myvar{u}_k^i,  \myvar{r}_k^i ) = 0.0001 \|\myvar{u}_k^i\|_2^2 + 10.0 \|\myvar{x}_k^i - \myvar{r}_k^i\|_2^2$ with the following state and input constraints: $-4.0  \leq \myvar{x}_k^i \leq 4.0 $, $-2.0  \leq \myvar{u}_k^i \leq 2.0 $ in \eqref{eq:DPC:objective}, and $N = 10$.

\begin{figure}
    \centering
    \begin{subfigure}{0.48\textwidth}
        \centering
        \includegraphics[width=\textwidth]{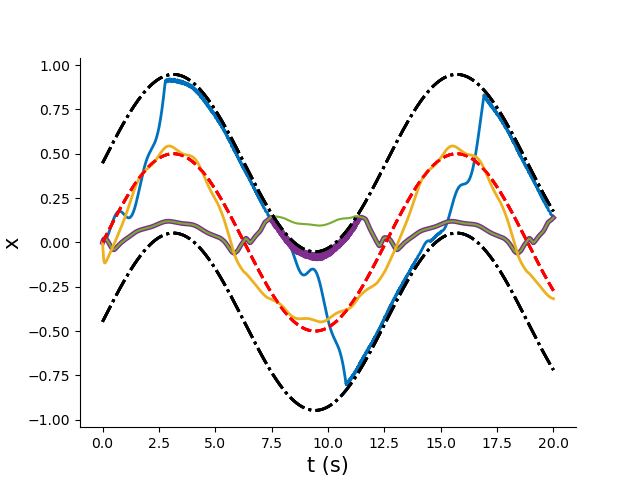}
        \caption{State trajectories with reference (red dashed curve) and safe set boundary, $\mysetbound{C}(t)$ (black dashed-dotted curves).}
        \label{fig:state_traj}
    \end{subfigure}
    
    \begin{subfigure}{0.48\textwidth}
        \centering
        \includegraphics[width=\textwidth]{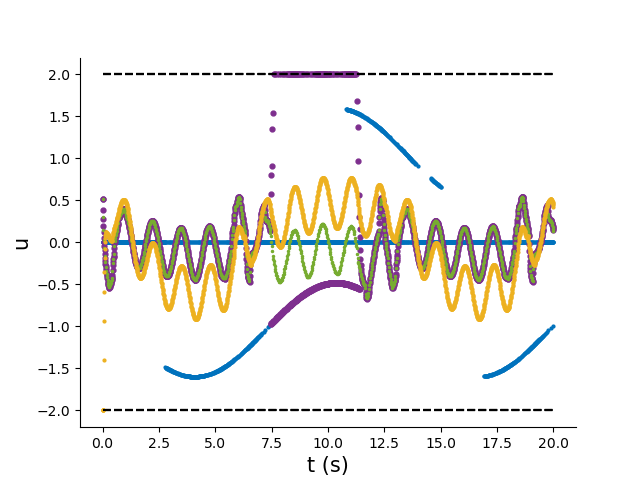}
        \caption{Input trajectories with input bounds (black dashed lines).}
        \label{fig:input_traj}
    \end{subfigure}
    
    \begin{subfigure}{0.48\textwidth}
        \centering
        \includegraphics[width=\textwidth]{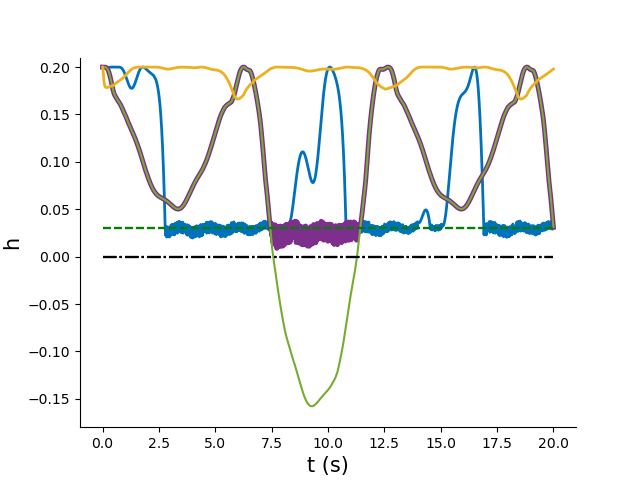}
        \caption{Barrier trajectories with $h = a$ (green dashed line).}
        \label{fig:barrier_traj}
    \end{subfigure}
    
    \caption{Comparison of trajectories for proposed control with no DPC (blue curve), proposed control with DPC (gold curve), DPC trained improperly (green curve), improperly trained DPC with proposed control (purple curve).}
    \label{fig:sim_results}
\end{figure}    

Figure \ref{fig:sim_results} compares several different implementations of the proposed controller and nominal DPC. First, the blue trajectories depict the proposed control \eqref{eq:safe DPC} with $\myvar{\pi}_W = 0$. The purpose of this simulation is to show how that the barrier controller alone is able to keep the system within the desired ultimate bound of the reference trajectory in the presence of a disturbance and sampling effects, while still respecting input constraints. Here the control can be seen to act as an event-triggering control where the QP is only solved when $h$ drops below $a$ as shown in Figure \ref{fig:barrier_traj}. The green trajectories depict nominal DPC ($\myvar{u} = \myvar{\pi}_W$), which was trained ``improperly", \bluetext{which means that the DPC control law was trained on a different reference of $\myvar{x}_r(t) = 0$, then implemented on the true reference}. Figures \ref{fig:state_traj} and \ref{fig:barrier_traj} show the trajectory leave the corridor and violate the desired ultimate bound condition. The purple trajectories depict the same improperly trained DPC controller implemented with the proposed control \eqref{eq:safe DPC}. This shows the proposed control correcting the learning controller to ensure safety. 

The final simulation depicted by the gold trajectory shows the ideal case where DPC was trained well \bluetext{i.e., trained with the true reference trajecotry,} and implemented in the proposed control \eqref{eq:safe DPC}. At no point during this simulation did the barrier function interfere with the DPC controller. This can be shown by seeing that the gold trajectory never drops below the $h = a$ line in Figure \ref{fig:barrier_traj}. This is the ideal performance of the proposed controller, where if the learned controller behaves well, the QP associated with \eqref{eq:safe DPC} is \emph{never} implemented. We note that the barrier function does not override the DPC controller even though at $t = 0$, DPC unexpectedly implements a maximum control input because the state lies in $\myset{C}(0)\setminus \myset{A}(0)$. As a result, the proposed control approximates an MPC control law without ever solving an optimization problem online. This fits with our theme that the barrier QP \eqref{eq:zcbf qp} should only be used as a backup in case unexpected outcomes of the learning controller or disturbance occur. \bluetext{This showcases another advantage over the previous SD-ZCBF, wherein even during the ideal case the SD-ZCBF method would require a QP to be solved at every sampling time. Clearly from these results, the proposed SD-ZCBFII significantly reduces the online computation requirements by only overriding the DPC controller when the system crosses the $h = a$ threshold.}

\bluetext{One limitation of the proposed solution is that for small $a$, the proposed control can exhibit bang-bang behavior due to the transition between the DPC control and barrier function QP. To mitigate this, $\|\myvar{u}_{k} - \myvar{u}_{k-1}\|_2^2$ can be added to the cost function of \eqref{eq:zcbf qp} to penalize sudden switches, as well as increasing $a$. Another limitation is the restriction to relative degree 1 barrier functions. Future work will extend the results to higher order systems. }

\section{Conclusion}

In this paper, we develop a provably safe, differentiable predictive control (DPC) law. DPC uses a differentiable programming-based policy gradient method to train a neural network to approximate an explicit MPC controller without the need for supervision from an expert controller. The proposed method entails using and developing a robust sampled-data barrier function to guarantee safety within a desired, time-varying safe set while implemented in an event-triggered fashion to reduce the computational resources used. This barrier function is then coupled with DPC to ensure the safety of the overall system. Simulation results demonstrate the viability of the proposed methodology. Future work will extend the approach to high-order barrier functions and extend DPC to nonlinear systems.

{
\bibliographystyle{IEEEtran}  
\bibliography{main}  
}

\end{document}